\documentclass[aip,reprint]{revtex4-1}
\usepackage{graphicx}
\draft % marks overfull lines with a black rule on the right
\usepackage{textcomp}

\begin{document}

\title{Insulating improper ferroelectric domain walls as robust barrier layer capacitors} 

\author{Lukas Puntigam}
\affiliation{Experimental Physics V, Center for Electronic Correlations and Magnetism, Institute of Physics, University of Augsburg, 86135 Augsburg, Germany}%
\author{Jan Schultheiß}%
\affiliation{Department of Materials Science and Engineering, Norwegian University of Science and Technology, 7043 Trondheim, Norway}
\author{Ana Strinic}
\affiliation{Experimental Physics V, Center for Electronic Correlations and Magnetism, Institute of Physics, University of Augsburg, 86135 Augsburg, Germany}%
\author{Zewu Yan}
\affiliation{Materials Sciences Division, Lawrence Berkeley National Laboratory, Berkeley, CA, 94720, USA}%
\affiliation{Department of Physics, ETH Zurich, 8093, Zürich, Switzerland}
\author{Edith Bourret}
\affiliation{Materials Sciences Division, Lawrence Berkeley National Laboratory, Berkeley, CA, 94720, USA}%
\author{Markus Altthaler}
\affiliation{Experimental Physics V, Center for Electronic Correlations and Magnetism, Institute of Physics, University of Augsburg, 86135 Augsburg, Germany}%
\author{Istvan Kezsmarki}
\affiliation{Experimental Physics V, Center for Electronic Correlations and Magnetism, Institute of Physics, University of Augsburg, 86135 Augsburg, Germany}%
\author{Donald M. Evans}%
\affiliation{Experimental Physics V, Center for Electronic Correlations and Magnetism, Institute of Physics, University of Augsburg, 86135 Augsburg, Germany}%
\affiliation{Department of Materials Science and Engineering, Norwegian University of Science and Technology, 7043 Trondheim, Norway}
\author{Dennis Meier}%
\affiliation{Department of Materials Science and Engineering, Norwegian University of Science and Technology, 7043 Trondheim, Norway}
\author{Stephan Krohns}
\affiliation{Experimental Physics V, Center for Electronic Correlations and Magnetism, Institute of Physics, University of Augsburg, 86135 Augsburg, Germany}%
\email[]{stephan.krohns@physik.uni-augsburg.de}

\date{\today}

\begin{abstract}
We report the dielectric properties of improper ferroelectric h-ErMnO$_3$. From the bulk characterisation we observe a temperature and frequency range with two distinct relaxation-like features, leading to high and even `colossal' values for the dielectric permittivity. One feature trivially originates from the formation of a Schottky barrier at the electrode-sample interface, whereas the second one relates to an internal barrier layer capacitance (BLC). The calculated volume fraction of the internal BLC (of 8\%) is in good agreement with the observed volume fraction of insulating domain walls (DWs). While it is established that insulating DWs can give rise to high dielectric constants, studies typically focused on proper ferroelectrics where electric fields can remove the DWs. In h-ErMnO$_3$, by contrast, the insulating DWs are topologically protected, facilitating operation under substantially higher electric fields. Our findings provide the basis for a conceptually new approach to engineer materials exhibiting colossal dielectric permittivities using domain walls in improper ferroelecctrics with potential applications in electroceramic capacitors.  

\end{abstract}


\maketitle %\maketitle must follow title, authors, abstract and \pacs

\section{INTRODUCTION} 
Materials exhibiting very high values in dielectric permittivity ($\varepsilon' > 10^3$) are often coined as ``colossal dielectric constant" (CDC) materials\cite{Lunkenheimer2010, Lunkenheimer2002}. They bear enormous potential for enhancing the capacitance, e.g. in multilayer ceramic or low-temperature co-fired capacitors \cite{Kishi2003, C9TC02921D, Li2020}. Typically, proper ferroelectric materials\cite{Toledano}, such as BaTiO$_3$\cite{Merz1949, Devon1949} or Pb(Zr$_x$,Ti$_{1-x}$)O$_3$\cite{Shirane1952}, are used as their dielectric permittivity exceeds $10^3$ at ambient temperatures and the loss tangent -- indicating the dielectric loss -- is rather low (tan$\delta < 10^{-2}$)\cite{C7TA05392D, C9TC02921D,Elissalde2001}. The frequency stability of these parameters within a certain temperature range with respect to the applied voltages are key quantities for technical applicability.

Another approach towards CDCs is to use thin layers with reduced conductivity -- so called barrier-layer capacitances (BLC) -- in bulk ceramics and single crystals. These BLCs can be internal layers, like insulating grain boundaries in polycrystalline ceramics\cite{Adams2002,Zhao2004,Frey1998}, or surface layers formed, e.g., due to the depletion zone of Schottky diodes arising at metal-semiconductor contacts\cite{Krohns2007,Krohns2009}. Both mechanisms are sensitive to variations in preparation (i.e., size of grains and conductivity of grain boundaries\cite{Zhao2004}) or the contact area of the Schottky barrier. BLCs appear as a step-like decrease in $\varepsilon'$ accompanied by a peak in $\varepsilon''$ in a frequency-dependent representation mimicking a classical `Debye-like' relaxation process (see \textcite{Lunkenheimer2010} and references therein for more details). The electrical heterogeneity
is responsible for the first relaxation-like feature in the dielectric properties, called Maxwell-Wagner relaxation.

Recently, a reduced conductivity at the DWs and a related BLC effect was observed in h-YMnO$_3$\cite{Ruff2017ConductivityYMnO3} suggesting a high dielectric constant $\varepsilon' > 200$. 
However, a systematic analysis that can confirm the connection between the high dielectric constant and DW driven BLCs remains illusive. 

Here, we provide a dielectric analysis of an h-ErMnO$_3$ single crystal for which the polarisation is parallel to the applied contact electrodes (in-plane). Two distinct relaxations are observed in this sample: the first leads to a high dielectric constant in the order of 300, and the second to a CDC of more than $5\times 10^3$. So far, mainly the dielectric properties of out-of-plane polarized samples were investigated, for which often only one relaxation was reported leading to CDC\cite{Holstad2018, Ruff2018APL, Schaab2018ElectricalWalls}. 
To disentangle the contribution of various BLCs arising in the sample and at the surface of the sample, we simulate the dielectric spectra by an equivalent circuit model, analogous to previous studies on the CDC prime-example, CaCu$_3$Ti$_4$O$_{12}$\cite{Krohns2008,Lunkenheimer2010,Lunkenheimer2002}. This approach, in combination with a distinct modification of the electrode contact area and the thickness of the sample, allows to distinguish BLCs arising from internal and surface effects. 
Furthermore, we use local probe analysis by piezo-force response (PFM) and conductive atomic force microscopy (cAFM) to determine electronic DWs properties at the sample surface and estimate the volume fraction of insulating DWs\cite{Meier2012,Schoenherr2018}. Our systematic analysis provides new insight into the dielectric properties of hexagonal manganites, corroborating that insulating DWs act as BLCs, playing a key role for the high or even colossal dielectric constants observed in this class of materials.

\section{EXPERIMENT} 

High-quality hexagonal h-ErMnO$_3$ single crystals were grown by the pressurized floating zone technique\cite{Yan2015}. The sample was cut into a disc (area = 2.38\,mm$^2$, thickness = 0.61\,mm) with the polar axis lying parallel to the surface, i.e. (110)-oriented. For dielectric spectroscopy we used a plate capacitor geometry, coating both top and bottom surfaces either with silver paint or sputtered gold. We performed the measurements using a Alpha Analyzer (Novocontrol, Montabaur, Germany), which covers the frequency range of 1\,Hz to 1\,MHz. This analysis was conducted in a closed-cycle refrigerator between 150\,K and 300\,K.

The microscopic data was recorded on the same sample at room-temperature using an NT-MDT (NTEGRA, Apeldoorn, Netherlands) atomic force microscope (AFM), using diamond tips (DDESP-10, Bruker, Billerica, MA, USA).
The sample was lapped with a 9\,\textmu m-grained Al$_2$O$_3$ water suspension and polished using silica slurry (Ultra-Sol® 2EX, Eminess Technologies, Scottsdale, AZ, USA) to produce a flat surface with a root mean square roughness (RMS) of about 1.65\,nm (determined over a $25\times25\,$\textmu m$^2$ area).
For domain imaging by PFM, an ac-excitation voltage of 10\,V was applied to the back electrode at a frequency of 61\,kHz while the tip was grounded. Local transport data was gained by cAFM with a dc-voltage of 2\,V applied to the back electrode.

\section{RESULTS \& DISCUSSION}

\subsection{Two relaxation-like dielectric features}

\begin{figure}[htb]
\centering
\includegraphics[width=\linewidth]{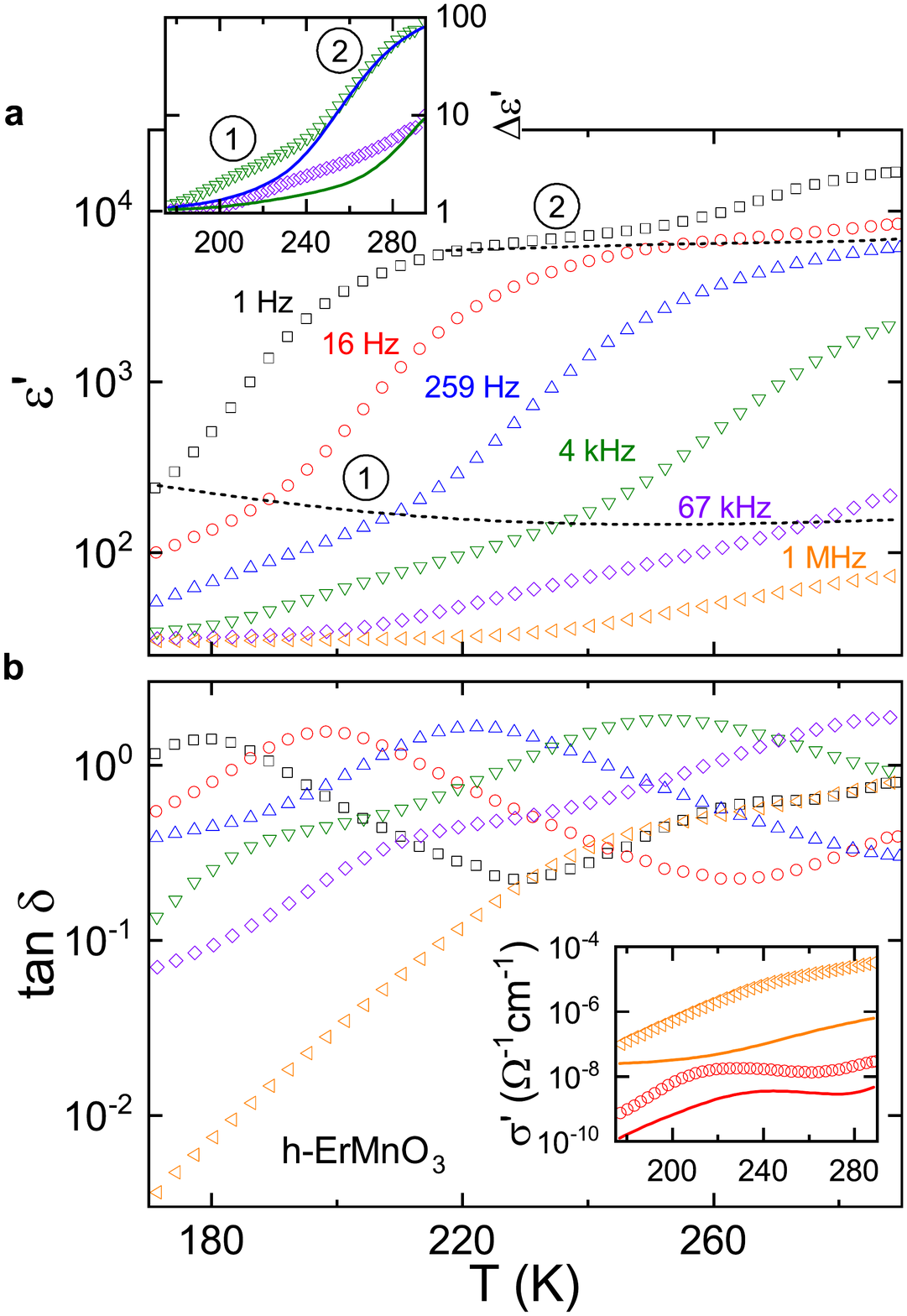}
\caption{Temperature-dependence of $\varepsilon'$ \textbf{a} and tan$\delta$ \textbf{b} of h-ErMnO$_{3}$ with in-plane polarisation at selected frequencies from 1\,Hz to 1\,MHz. The inset in \textbf{a} compares $\Delta\varepsilon' = \varepsilon'/\varepsilon_{\infty}$ of the in-plane (open symbols) and out-of-plane (lines) polarisation of samples from the same batch for two different frequencies (colours refer to the respective frequencies in \textbf{a}). The data for the out-of-plane sample were taken from \textcite{Ruff2018APL}. Numbers (\textcircled{\footnotesize{1}} and \textcircled{\footnotesize{2}}) indicate the relaxation-like features (dashed lines give a guide to the eyes for the temperature-dependency of these features). The inset in \textbf{b} compares the temperature dependent conductivity $\sigma'$ for these samples for two frequencies (again, colours refer to the respective frequencies in \textbf{a}).
}
\label{fig1}
\end{figure}

Figure \ref{fig1} shows the temperature dependent dielectric constant $\varepsilon'$ \textbf{a} and loss-tangent tan$\delta$ \textbf{b} for various frequencies from 1\,Hz to 1\,MHz. The electrodes where made with silver paint. For almost all frequencies $\varepsilon'$ exhibits a distinct two-step increase from about 30 to 200 -- 400 and further to $\varepsilon' > 5\times 10^3$. These steps in $\varepsilon'$ are accompanied by a peak in tan$\delta$, e.g., for the 4\,kHz curve at about 185\,K and 255\,K, respectively. This behaviour corresponds to a relaxation process in the temperature-dependent representation. Such prominent relaxation-like features are well known in oxide materials\cite{Lunkenheimer2010}, often originating from BLCs, e.g., Schottky-diodes forming a depletion-zone that acts as a thin insulating layer. 
Further, the rather high values of the loss tangent (tan$\delta > 0.01$) corroborate the framework of BLC mechanisms responsible for the dielectric features. 

Recently, it was reported that in h-YMnO$_3$ and h-ErMnO$_3$ the Schottky-barriers give rise to CDCs\cite{Schaab2018ElectricalWalls,Ruff2017ConductivityYMnO3,Holstad2018,Ruff2018APL}. Similar to this work (Fig.\,\ref{fig1}), two distinct relaxation-like features on h-YMnO$_3$ were measured\cite{Ruff2017ConductivityYMnO3}; one attributed to an internal BLC mechanism, possibly originating from insulating DWs. In contrast to the previously published data gained on samples with out-of-plane polarisation, the measurements presented in Fig.\,\ref{fig1} show well separated features, facilitating a more detailed analysis. We illustrate this difference with the inset in Fig.\,\ref{fig1}\,\textbf{a}, comparing the change in dielectric constant $\Delta\varepsilon'$ for the present in-plane oriented sample to an out-of-plane oriented one, published in \textcite{Ruff2018APL}. The CDC feature \textcircled{\footnotesize{2}} is the same, while the high dielectric constant feature \textcircled{\footnotesize{1}} appears as a distinct increase only for the in-plane sample. It is important to note, that for this comparison of $\varepsilon'$ we used different frequencies for the in-plane (259\,Hz and 4\,kHz) and out-of-plane (4\,kHz and 67\,kHz) orientation. For BLC-driven mechanisms the bulk dc-conductivity has a strong impact on the temperature and frequency range, where this feature dominates the dielectric properties\cite{Lunkenheimer2010}. 
The inset of Fig.\,\ref{fig1}\,\textbf{b} shows $\sigma'$ for two representative frequencies, indicating a significant decrease of the conductivity of the out-of-plane oriented sample corroborating the above mentioned shift in the BLC driven feature. This is confirmed by the frequency-dependent dielectric analysis of h-(Er$_{0.99}$Ca$_{0.01}$)MnO$_3$ shown in Fig.\,S1. 
This is a first hint of an anisotropic BLC feature, which is either based on the change in dc-conductivity or might be due to differences in DW density or conductance. For the latter, it is already well established via local probe measurements, that the conductance of the DWs strongly depend on the orientation of the polarisation\cite{Meier2012,Mosberg2019}.

\subsection{Quantifying the DW density}

\begin{figure}[htb]
\centering
\includegraphics[width=\linewidth]{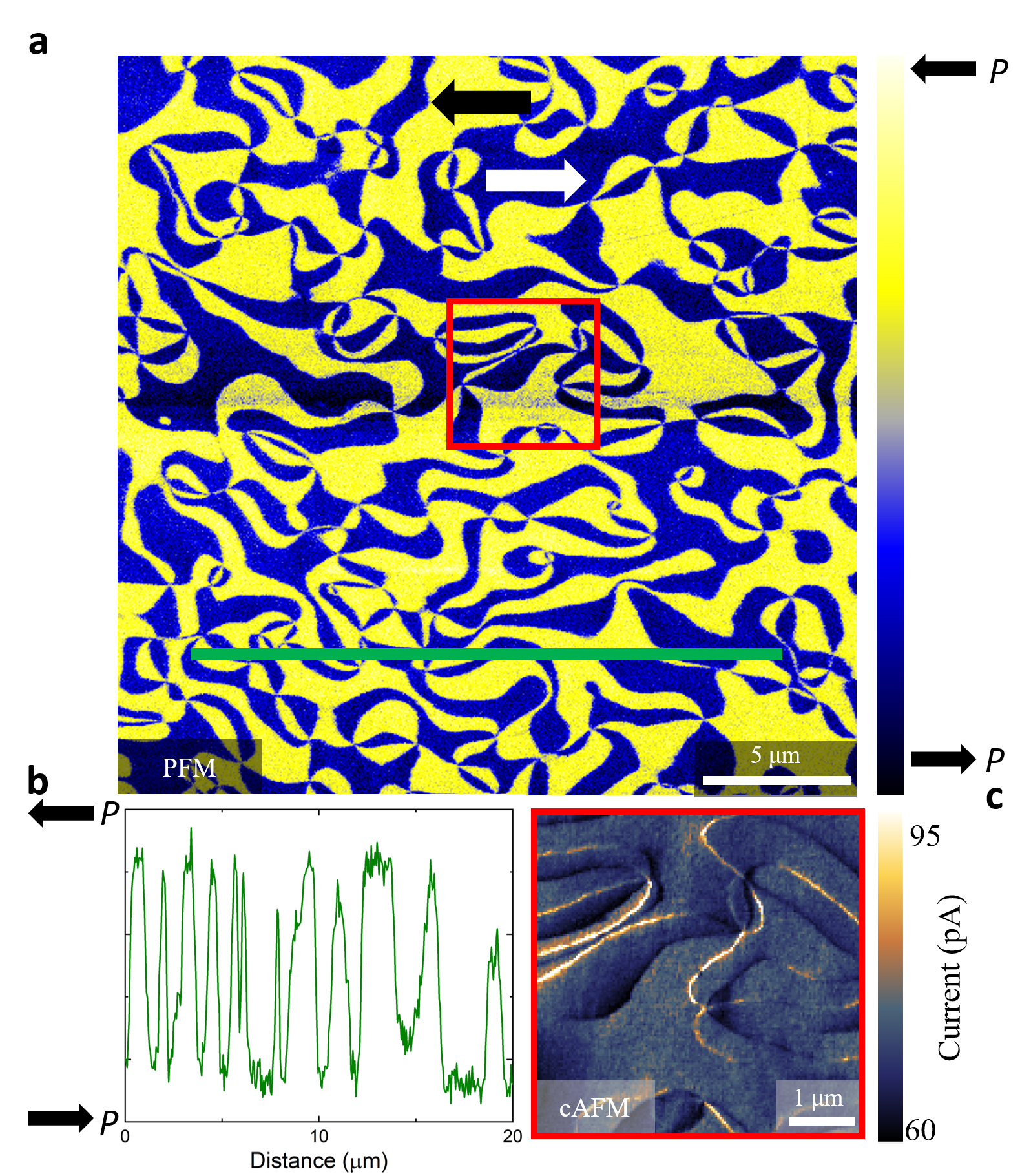}
\caption{\textbf{a} Calibrated in-plane PFM scan where the yellow (blue) represents ferroelectric domains pointing to the left (right). These form the characteristic domain pattern present in hexagonal manganite single crystals, where the ferroelectric $180^{\circ}$ domains come together at sixfold vertex points. \textbf{b} A representative line-profile from \textbf{a} where high values are from domains with \textit{$\leftarrow$P} and low values are from \textit{$\rightarrow$P} domains, used to get an approximate density of DWs per length. \textbf{c} A cAFM image from the area marked be the red box in \textbf{a}. Light colors indicated areas of enhanced conductance (tail-to-tail walls) while dark areas indicate areas of lower conductance (head-to-head walls).}
\label{fig2}
\end{figure}

We use PFM and cAFM to estimate the density of DWs, which provide a possible origin for the observed high dielectric constant feature.
To approximate the density of these
DWs in our h-ErMnO$_3$ crystal, we map the domain distribution at the sample surface using PFM (in-plane contrast) as presented in Fig.\,\ref{fig2}\,\textbf{a}. The PFM scan reveals the typical domain distribution, characteristic for hexagonal manganites\cite{Choi2010,Jungk2010,Safrankova1967}. 
To determine the domain / DW density, we evaluate multiple test lines (one line is shown in Fig.\,\ref{fig2}\,\textbf{b} and further lines in Fig.\,S3) applying the procedure outlined in \textcite{Hubert1998}.
Measurements by cAFM (Fig.\,\ref{fig2}\,\textbf{c}) confirm the presence of DWs with enhanced (tail-to-tail) and reduced (head-to-head) conductance\cite{Meier2012}.
From this analysis we find $1 \pm 0.1$ DWs per \textmu m with enhanced or reduced conductance in comparison to the bulk. The same evaluation was also performed for h-(Er$_{0.99}$Ca$_{0.01}$)MnO$_3$ depicted in Fig.\,S2, providing a similar domain / DW fraction.  

\subsection{Dielectric features due to barrier layers}

\begin{figure}[htb]
\centering
\includegraphics[width=\linewidth]{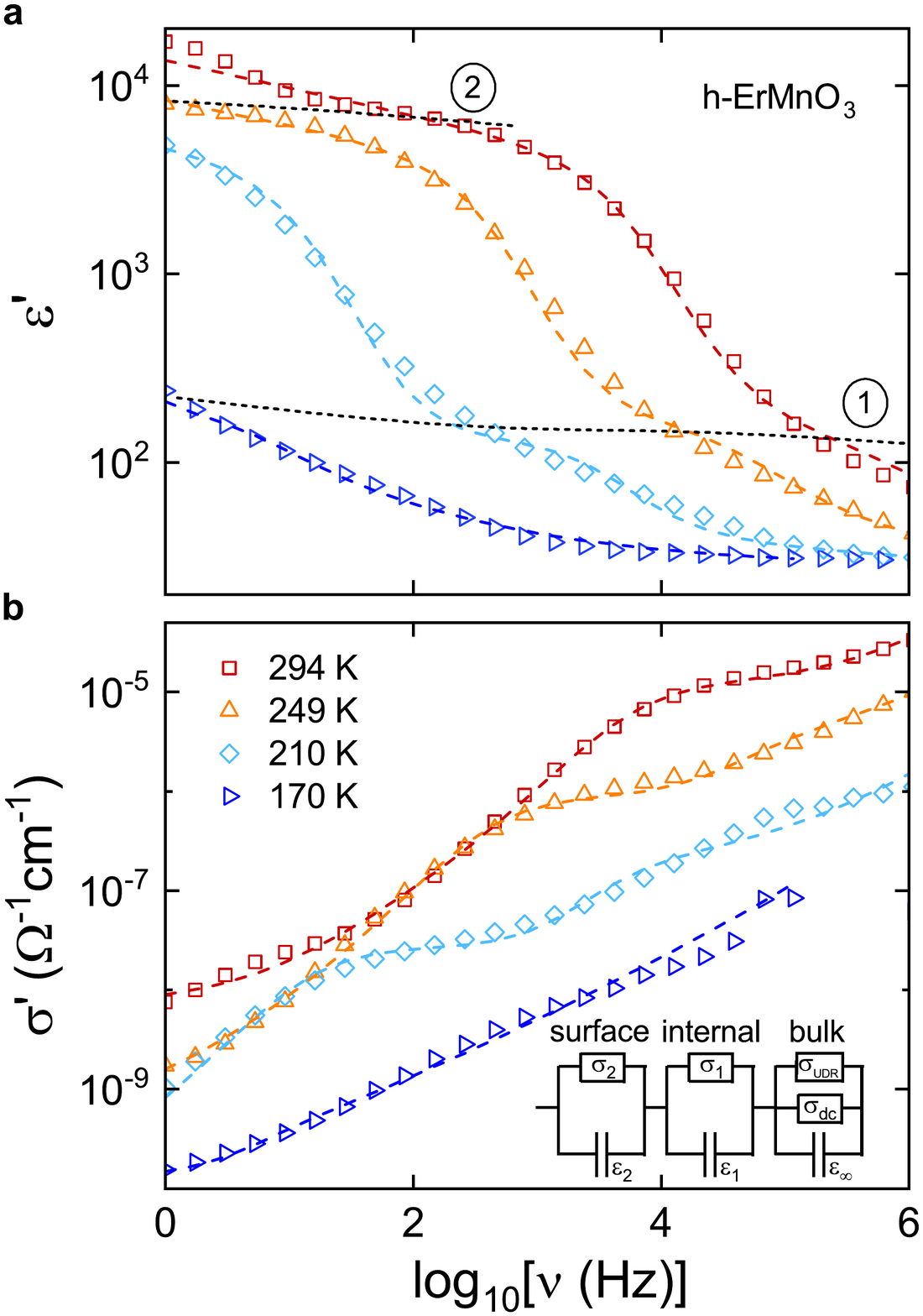}
\caption{Frequency-dependence of $\varepsilon$' \textbf{a} and $\sigma$' \textbf{b} of h-ErMnO$_{3}$ with in-plane polarisation at various temperatures (again numbers \textcircled{\footnotesize{1}} and \textcircled{\footnotesize{2}} as well as the blacked dashed lines indicate the relaxation-like features). The coloured dashed lines represent fits for the measured data, obtained by considering the equivalent circuit model, sketched in the corner of \textbf{b}. The equivalent circuit model consists of three $RC$-circuits connected in series.}
\label{fig3}
\end{figure}

To disentangle surface and internal contributions, a frequency-dependent analysis of the dielectric response is required, which is shown in Fig.\,\ref{fig3}. The frequency-dependent dielectric permittivity (Fig.\,\ref{fig3}\,\textbf{a}) exhibits two distinct relaxations for selected temperatures varying from 170\,K to 294\,K. The first one in the 210\,K-curve evolves at 1\,kHz indicated by a step-like increase in $\varepsilon'$ from $\sim20$ to $\sim300$. The lower $\varepsilon'$-value at high frequencies denotes the contribution of ionic and electronic polarizability to the so-called intrinsic $\varepsilon_{\infty}$, confirming literature values in the order of $20 - 40$\cite{Holstad2018,Ruff2018APL}. 
The upper plateau of the second step for the 210\,K-curve begins at $\nu < 100$\,Hz and settles at an $\varepsilon'$ value in the order of $5 \times 10^3$. Both relaxations are accompanied by steps in $\sigma' (\nu)$. The plateaus in $\sigma'$ indicate roughly the dc-conductivity of the BLCs and the bulk. However, the dc-conductivity of step two -- most likely the Schottky barrier -- is almost shifted out of the measured frequency range. The curvature of $\sigma'$(294\,K) for $\nu < 10$\,Hz indicates the onset of this dc-plateau of approximately $\sigma_{dc} \approx 3\times 10^{-9}$\,$\Omega^{-1}$cm$^{-1}$. The $\sigma_{dc}$-plateau of the first BLC feature emerges, e.g., for the 210\,K-curve between 100 -- 1000\,Hz at $2\times 10^{-8} $\,$\Omega^{-1}$cm$^{-1}$. Finally, at higher frequencies, e.g., $\nu > 10$\,kHz for 210\,K-curve, $\sigma_{dc}$ of the bulk evolves, which is for low temperatures superimposed by a contribution of an universal dielectric response (UDR) feature\cite{Jonscher1977}, giving rise to a frequency-dependent increase in the overall conductivity.

In oxide materials there exist different descriptions for these observed dielectric relaxation-like features: (i) hopping conductivity, (ii) displacements of DWs in excitation fields far below the coercive field, (iii) charged DWs acting as conductive inclusions\cite{Esin2017}, (iv) Maxwell-Wagner type electrical inhomogeneities or (v) Schottky-barriers at the electrode contacts. Hopping conductivity in disordered systems can be responsible for a continuous increase in $\varepsilon'$ at low frequencies (typical $\nu < 1$\,Hz)\cite{Lunkenheimer2010}. In the present study, we observe distinct relaxation-like features strongly shifting with temperature, excluding mechanism (i). Further, the DWs are strongly pinned by the vortex structure of the hexagonal manganites and the coercive-fields for polarisation reversal in h-ErMnO$_3$ are typically $> 3$\,kV/mm\cite{Choi2010,Ruff2018APL}, which is orders in magnitude higher than the used voltage of 1\,V/mm for the dielectric spectroscopy. Therefore, irreversible DW motion does not contribute to the dielectric permittivity\cite{Lewis1959,Hall1999}. In addition, the dielectric relaxation related to reversible domain wall motion typically occurs in the GHz range\cite{Kittel1951} and is temperature independent\cite{Arlt1994}. The strong temperature-dependence of the dielectric relaxation observed in our samples (Fig.\,\ref{fig1}) also allows to exclude reversible domain wall motion as an origin. 

Due to these robust domain structure, mechanism (ii) seems to be very unlikely. Interestingly, explanation model (iii) requires charged conductive DWs, which are discussed to enhance the dielectric permittivity in (K,Na)NbO$_3$-based ferroelectrics\cite{Esin2017}. More specific, a higher density of DWs leads to an increase in dielectric permittivity. Charged DWs with enhanced conductivity are also present in our samples (c.f. Fig.\,\ref{fig2}\,\textbf{c}). However, in contrast to the above mentioned ferroelectric, we find the contrary behaviour of the relation of DW density to the dielectric permittivity. Here, $\varepsilon'$ decreases with increasing DW density\cite{Ruff2017ConductivityYMnO3}, excluding charged conductive DWs as origin for the observed BLC feature. Thus, we focus in the following only on the latter two mechanisms of internal (iv) and surface (v) BLCs. For the present study we investigated a single crystal, for which we can further exclude a BLC mechanism due to insulating grain boundaries, as observed in oxide ceramics\cite{Adams2002,Lunkenheimer2010,Moulson}.

The dashed lines in Fig.\,\ref{fig3} represent fits using the equivalent circuit model depicted in the inset of Fig.\,\ref{fig3}\,\textbf{b}. This model uses the approach of Maxwell and Wagner\cite{Wagner1914,Maxwell1873}, for which volume fractions of the overall sample with certain capacitances and conductivities can be described by discrete $RC$-circuits connected in series. In a nutshell, we deploy $RC$-circuits for step two (surface BLC) and step one (internal BLC) in series to the bulk properties. For the latter $RC$-circuit (bulk) we use an additional frequency-dependent resistor accounting for the power-law contribution of UDR to $\sigma'$\cite{Jonscher1977}. From these fits parameters of the dielectric constants and dc-conductivities ($\varepsilon_{\infty}$', $\varepsilon_{1}$', $\varepsilon_{2}$', and temperature-dependent values for $\sigma_{dc}, \sigma_1'$ and $\sigma_2'$) are gained as listed in Table S1. The fitting parameters confirm the temperature-dependent evolution of the dielectric properties of the BLC contributions and the semiconducting behaviour of the bulk.

While fitting with an equivalent circuit model allows for analyzing and disentangling single BLC contributions, it cannot provide information about the underlying mechanism. The formation of a Schottky barrier at the interface of the metal electrode and the semiconducting bulk leads to a depletion layer that acts as a thin capacitor at the sample surface. A proven approach\cite{Krohns2007,Lunkenheimer2010} to establish such Schottky BLCs is to measure the dielectric properties using electrodes of different wettings, e.g. painted \textit{vs}. sputtered electrodes: for Schottky BLCs an increased wetting will provide an enhanced CDC. Internal BLCs and the intrinsic dielectric properties, however, should not be affected by this change. Another way to proof a surface BLC is to reduce the sample thickness, as both $\varepsilon'$ and $\sigma'$ are geometry independent values, so the dielectric response should not be affected.

\begin{table*}[htb]
\caption{Calculated and measured contribution of the DWs to the dielectric response for h-\textit{R}MnO$_3$ (\textit{R} = Y, Er) with in-plane (IP) and out-of-plane (OOP) polarisation. The values are estimated from the dielectric spectroscopy (Figs. \ref{fig3} \& S1) and local-probe analyses (Figs. \ref{fig2}, S2 \& S3).}
\label{tab1}
\begin{tabular}{l|c|c|c|c|c|c}
sample & dir. of P & $\varepsilon_{\infty}$ & $n_{DW}$ [$1/$\textmu\text{m}] & $V_{DW}$ [\%] & estim. $\varepsilon_{1}$ & meas. $\varepsilon_{1}$ \\ %$\sigma_{dc}($300\,K$)[\Omega^{-1}$cm$^{-1}]$\\
\hline
h-ErMnO$_3$ (Figs. \ref{fig2}, \ref{fig3} \& S3) & IP & 32 $(\pm 3)$ & 0.5 $(\pm 0.05)$ & 7.5 $(\pm 0.75)$ & 400 $(\pm 60)$ & 250 $(\pm 50)$  \\
%h-ErMnO$_3$ (Figs. S1 \& S3) & IP & 25 & 0.5 & 8 & $310$ & $260$\\ % $\gg 1 \times 10^{-4}$ \\
h-(Er$_{0.99},$Ca$_{0.01}$)MnO$_3$ (Figs. S1 \& S2) & IP & 18 & 0.46 & 7 & $260$ & $220$ \\ %$\gg 1 times 10^{-4}$\\ 
\hline
h-YMnO$_3$ sample 1 (data from \textcite{Ruff2017ConductivityYMnO3}) & OOP & 20 & 0.17 & 2.5 & $780$ & $670$ \\% $1\times 10^{-7}$ \\
h-YMnO$_3$ sample 2 (data from  \textcite{Ruff2017ConductivityYMnO3}) & OOP & 20 & 2.5 & 37.5 & 54 & 40 \\ %$3 \times 10^{-7}$ \\
\end{tabular}
\end{table*}

\begin{figure}[htb]
\centering
\includegraphics[width=\linewidth]{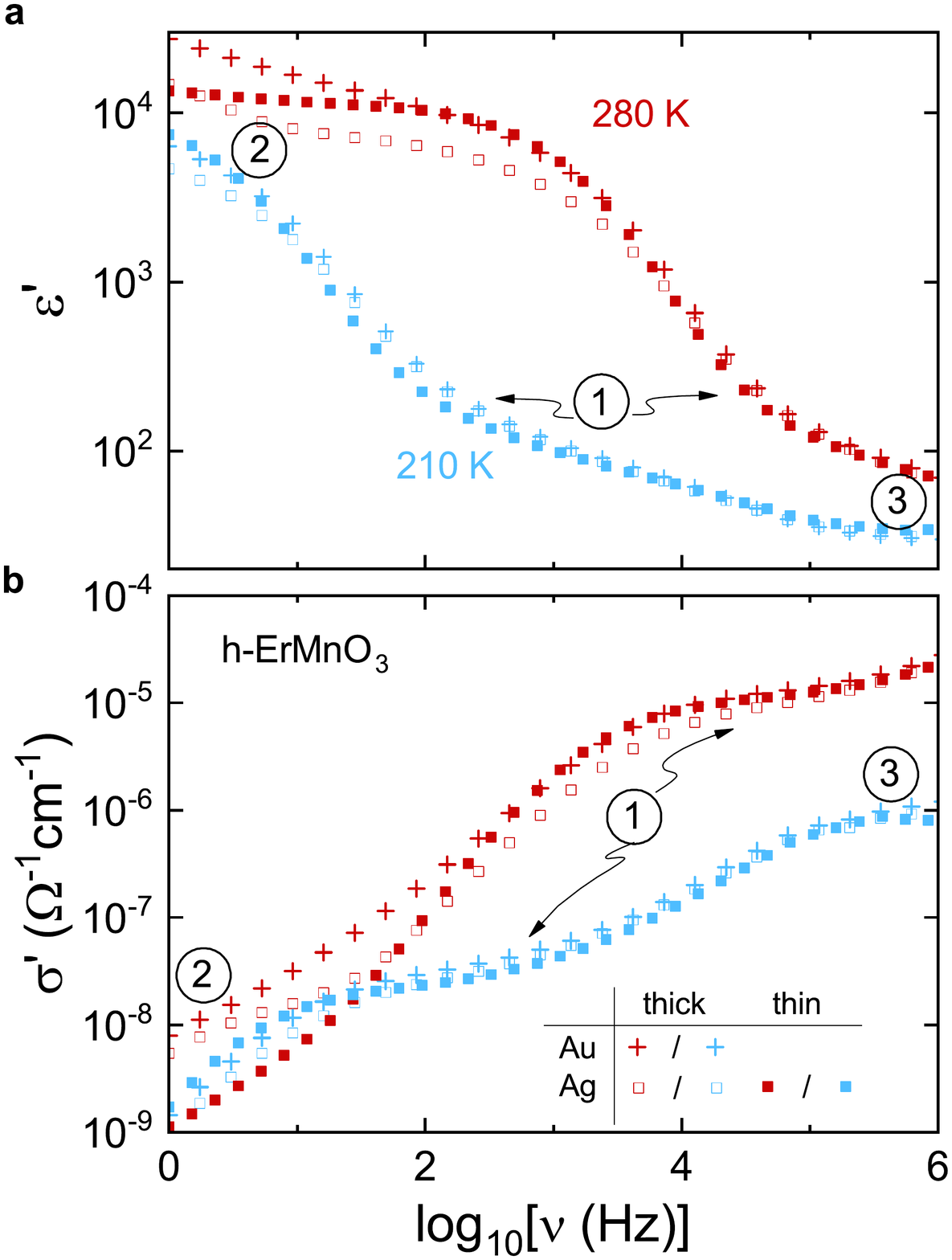}
\caption{Frequency-dependence of $\varepsilon$' \textbf{a} and $\sigma$' \textbf{b} for two representative temperatures. In this figure the measurement data for sputtered gold and silver paint as contacting material are shown and, for the latter, also the thickness dependence ($d_{\text{thick}}$ = 1\,mm and $d_{\text{thin}}$ = 0.6\,mm). Numbers indicate the dielectric features: \textcircled{\footnotesize{1}}$\rightarrow$ BLC one (internal), \textcircled{\footnotesize{2}}$\rightarrow$ BLC two (surface), 
\textcircled{\footnotesize{3}}$\rightarrow$ bulk properties.}
\label{fig4}
\end{figure}

Figure\,\ref{fig4} shows frequency-dependent $\varepsilon'$ \textbf{a} and $\sigma'$ \textbf{b} for both strategies indicated by different symbols (open symbols $\rightarrow$ silverpaint \& $d_{\text{thick}}$, closed symbols $\rightarrow$ silverpaint \& $d_{\text{thin}}$, crosses $\rightarrow$ sputtered gold \& $d_{\text{thick}}$). Analogous to Fig.\,\ref{fig3} both relaxations appear for the different surface treatments. 
Importantly, we find that the intrinsic bulk properties \textcircled{\footnotesize{3}}, as well as the first relaxation \textcircled{\footnotesize{1}} step, are -- as expected for the bulk properties and an internal BLC -- independent of the electrode material and the thickness of the sample. In contrast, the enhanced wetting of the sputtered electrode increases the upper plateau of $\varepsilon'$ by a factor of about two. Furthermore, reducing the sample thickness gives rise to an increase in the CDC feature. 
This leads us to the conclusion that the second relaxation \textcircled{\footnotesize{2}} is due to the formation of an insulating depletion layer due to a Schottky barrier at the sample surface. Critical, and in contrast, the first relaxation is not affected by these changes to the surface, corroborating the hypothesis of an internal BLC mechanism related to insulating DWs\cite{Ruff2017ConductivityYMnO3}.

\subsection{Insulating DW barrier layer capacitors}

From the local probe analysis of the sample surface (Fig.\,\ref{fig2}), in combination with bulk dielectric measurements, we calculate an approximate volume fraction (in \%) of the insulating DWs in our sample: $V_{DW} = n_{DW} d_{DW}$, where $n_{DW}$ denotes the number of insulating DWs per \textmu m and $d_{DW}$ the electronic thickness of the DWs referred to as electrical dressing in \textcite{Meier2012}. Note, this electronic width is much larger than the structural width of about 5\,Å\cite{Holtz2017}, reaching values in the order of 100 to 150\,nm, which was related to a delocalisation of charge carriers\cite{Meier2012}.
As the contributions of the ionic and electronic polarizability to $\varepsilon_{\infty}$ for both bulk and DWs are almost the same\cite{Schaab2018ElectricalWalls}, we can estimate the value of the dielectric constant $\varepsilon_{1}$. Within the framework of the Maxwell-Wagner model, $\varepsilon_{1}$ is given by the relation: $\varepsilon_{1} \approx \varepsilon_{\infty}/V_{DW}$.

In table \ref{tab1}, the calculated values for the dielectric response of different sample are compared.
The estimated and measured values for all samples listed in table \ref{tab1} are almost in the same order, providing a strong indication for a BLC-DW based mechanism. However, we note that a smeared-out relaxation due to a distribution of relaxation times of the BLC, an inhomogeneous polar pattern, as well as possible variations of the intrinsic dielectric constant cannot be excluded, adding to the uncertainty of the presented values in table \ref{tab1}. 

Finally, we address the question, why in-plane polarized samples seem to be suited better than out-of-plane polarized specimens for detecting BLC-DW features. This can be explained employing the Maxwell-Wagner model: The relaxation-time $\tau_{BLC-DW}$ of the BLC-DW $RC$-circuit connected in series to the bulk is: $\tau_{BLC-DW} \propto \varepsilon'_{1}/\sigma_{dc}$. Thus, $\varepsilon'$ -- mainly based on the number of insulating DWs and their effective thickness -- and the temperature-dependent bulk $\sigma_{dc}$ are strongly affecting the frequency range in which the relaxation occurs. Especially, for the samples h-ErMnO$_3$ and h-(Er$_{0.99}$,Ca$_{0.01}$)MnO$_3$, we measured a strongly anisotropic dc-conductivity, which is enhanced for the samples with in-plane polarisation by up to a factor of 100 -- depending on the temperature -- compared to the samples with out-of-plane polarisation (Figs.\,S1 \& S2). An enhanced dc-conductivity results in a decrease in $\tau_{BLC-DW}$ and, thus in an increase in the corresponding frequency $\nu_{BLC-DW} = 1/(2\pi\tau_{BLC-DW})$. According to this, the anisotropy in the bulk $\sigma_{dc}$ seems to be the most likely reason, which allows us to disentangle the contributions of surface and internal BLCs, especially in the case of samples of h-ErMnO$_3$ with in-plane polarisation.

\section{SUMMARY}

In this study, we investigate the contribution of insulating DWs in h-ErMnO$_3$ to the overall dielectric response up to the MHz regime. Depending on the temperature range of the dielectric measurements, two distinct relaxation-like processes are revealed, especially for samples with in-plane polarisation. One of these relaxation-like features originates from a Schottky barrier, which we proof by target-oriented manipulation of the sample-electrode interface. For the other feature we conclude, corroborated by PFM and cAFM measurements, that an internal barrier layer is formed by insulating DWs. To proof this hypothesis, we first use an equivalent-circuit model to quantify the bulk dielectric properties of this internal barrier layer. Secondly, we compare these values to a barrier layer calculated with the density and the electronic thickness of insulating DWs measured by PFM and cAFM. Based on this data, we confirm that internal barrier layer capacitors are formed at insulating DWs, which is corroborated by the comparison of different h-YMnO$_3$ and h-ErMnO$_3$ samples.  

As both the density\cite{PhysRevX.7.041014, PhysRevX.2.041022} 
and electrical characteristic of the insulating DWs\cite{Holstad2018, Hassanpour_2016} can be tuned, the engineering the macroscopic dielectric response is feasible. This may pave the way to generate high and even colossal dielectric constants by robust internal barrier layers in h-ErMnO$_3$ and h-YMnO$_3$, which makes these improper ferroelectrics promising for the use as dielectrics in multilayer ceramic capacitors.

\section*{Author contributions}
S.K. initiated and coordinated the project. L.P. and A.S. conducted the dielectric experiments, and L.P., I.K. and S.K. analyzed the data. J.S., L.P., M.A. and D.E. performed and analyzed the local-probe experiments supervised by D.M. The single crystals were prepared by E.B. and Z.Y. All authors participated in the discussion and interpretation of the results. D.E., L.P., J.S., D.M. and S.K. wrote the manuscript.

\section*{Acknowledgments}
J.S. acknowledges the support from the Alexander von Humboldt Foundation through the Feodor-Lynen fellowship. D.M. acknowledges support by NTNU through the Onsager Fellowship Program and the Outstanding Academic Fellows Program, and funding from the European Research Council (ERC) under the European Union`s Horizon 2020 Research and Innovation Programme (Grant Agreement No. 86691). S.K., M.A., A.S., L.P., and I.K. acknowledge the funding of the German Science foundation via the Collaborative Research Center TRR80.

\section*{Data Availability}
The data that support the findings of this study are available from the corresponding author upon reasonable request.

\end{document}

\title{Insulating improper ferroelectric domain walls as robust barrier layer capacitors} %Title of paper



\author{Lukas Puntigam}
\affiliation{Experimental Physics V, Center for Electronic Correlations and Magnetism, Institute of Physics, University of Augsburg, 86135 Augsburg, Germany}%
\author{Jan Schultheiß}%
\affiliation{Department of Materials Science and Engineering, Norwegian University of Science and Technology, 7043 Trondheim, Norway}
\author{Ana Strinic}
\affiliation{Experimental Physics V, Center for Electronic Correlations and Magnetism, Institute of Physics, University of Augsburg, 86135 Augsburg, Germany}%
\author{Zewu Yan}
\affiliation{Materials Sciences Division, Lawrence Berkeley National Laboratory, Berkeley, CA, 94720, USA}%
\affiliation{Department of Physics, ETH Zurich, 8093, Zürich, Switzerland}
\author{Edith Bourret}
\affiliation{Materials Sciences Division, Lawrence Berkeley National Laboratory, Berkeley, CA, 94720, USA}%
\author{Markus Altthaler}
\affiliation{Experimental Physics V, Center for Electronic Correlations and Magnetism, Institute of Physics, University of Augsburg, 86135 Augsburg, Germany}%
\author{Istvan Kezsmarki}
\affiliation{Experimental Physics V, Center for Electronic Correlations and Magnetism, Institute of Physics, University of Augsburg, 86135 Augsburg, Germany}%
\author{Donald M. Evans}%
\affiliation{Experimental Physics V, Center for Electronic Correlations and Magnetism, Institute of Physics, University of Augsburg, 86135 Augsburg, Germany}%
\affiliation{Department of Materials Science and Engineering, Norwegian University of Science and Technology, 7043 Trondheim, Norway}
\author{Dennis Meier}%
\affiliation{Department of Materials Science and Engineering, Norwegian University of Science and Technology, 7043 Trondheim, Norway}
\author{Stephan Krohns}
\affiliation{Experimental Physics V, Center for Electronic Correlations and Magnetism, Institute of Physics, University of Augsburg, 86135 Augsburg, Germany}%
\email[]{stephan.krohns@physik.uni-augsburg.de}
%\thanks{}


\maketitle %\maketitle must follow title, authors, abstract and \pacs

\subsection{Parameters of equivalent circuit analysis}
Table I lists the fit parameters of the equivalent-circuit analysis performed for the data partially presented in Fig.\,3. The equivalent circuit, depicted as inset in Fig.\,3\,\textbf{b}, consists of three $RC$-elements, one for the bulk and two for barrier-layer contributions. The bulk $RC$-circuit also takes into account contributions of a frequency-dependent conductivity, the UDR behaviour\cite{Jonscher1977}. Details of the fitting routine can be found in \textcite{Lunkenheimer2010} and \textcite{Ruff2017ConductivityYMnO3}. Here, we list only the values for the dielectric properties. The additional quantities within the fitting model, i.e., the broadening-parameters $\alpha$, which comprises the distribution of relaxation times, and the UDR behaviour remain almost constant for the selected temperatures. 

\begin{table*}[htb]
    \caption{Fit parameters of the equivalent circuit analysis of the frequency-dependent dielectric properties of h-ErMnO$_3$}
    \centering
\begin{tabular}{c|c|c|c|c|c|c}

Temperature (K) &   $\varepsilon_{\infty}$  &   $\varepsilon_1$ &   $\varepsilon_2$ &   $\sigma_{dc}$ ($\Omega^{-1}\text{cm}^{-1}$)  &   $\sigma_{1}$ ($\Omega^{-1}\text{cm}^{-1}$) &   $\sigma_{2}$ ($\Omega^{-1}\text{cm}^{-1}$)\\
    \hline
290 &   35   &  200 &   14000  &   $1.45\cdot10^{-4}$   &   $1.56\cdot10^{-5}$  &   $7.29\cdot10^{-9}$\\

270 &   35   &  230 &   11941  &   $2.90\cdot10^{-5}$   &   $5.50\cdot10^{-6}$  &   $3.3\cdot10^{-9}$\\

250 &   35   &  231 &   11585  &   $7.10\cdot10^{-6}$   &   $1.14\cdot10^{-6}$  &   $6.13\cdot10^{-10}$\\

230 &   35   &  200 &   13000  &   $2.37\cdot10^{-6}$   &   $2.22\cdot10^{-7}$  &   $6.21\cdot10^{-11}$\\

210 &   35   &  200 &   13000  &   $2.34\cdot10^{-7}$   &   $3.26\cdot10^{-8}$  &   $2.66\cdot10^{-11}$\\

190 &   35.3   &  300 &   13000  &   $9.84\cdot10^{-9}$   &   $2.85\cdot10^{-9}$  &   $2.22\cdot10^{-14}$\\

170 &   35.8   &  300 &   13000  &   $7.53\cdot10^{-10}$   &   $2.15\cdot10^{-10}$  &   $3.66\cdot10^{-15}$\\
\hline
\end{tabular}
\end{table*}

\subsection{Dielectric and local probe measurements of h-(Er$_{0.99}$,Ca$_{0.01}$)MnO$_3$} 

\begin{figure}[htb]
    \centering
    \includegraphics[width=0.7\linewidth]{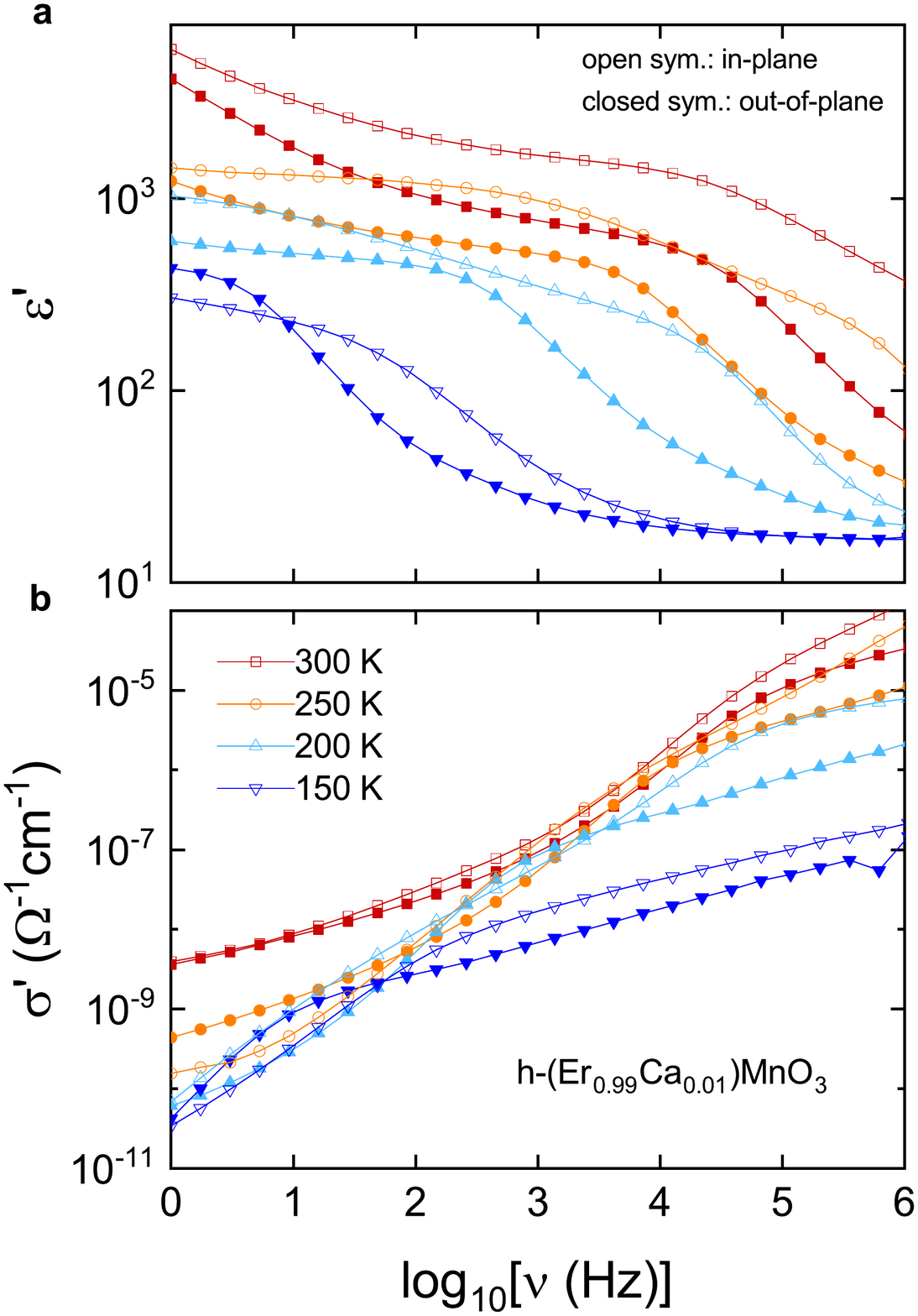}
    \caption{Frequency-dependent dielectric constant $\varepsilon$' \textbf{a} and conductivity $\sigma$' \textbf{b} h-(Er$_{0.99}$,Ca$_{0.01}$)MnO$_3$ sample with out-of-plane (closed symbols) and in-plane (open symbols) polarisation at various temperatures.}
    \label{SFigure1}
\end{figure}

Beside the undoped h-ErMnO$_3$ we performed the bulk dielectric and local-probe analyses on two 1\% Calcium doped ErMnO$_3$ samples of the same batch with out-of-plane and in-plane polarisation. Details of sample preparation can be found in \textcite{Hassanpour_2016}.
Figure S\ref{SFigure1} shows the frequency dependent dielectric properties for samples with out-of-plane (closed symbols) and in-plane (open symbols) polarisation. Again, in analogy to the discussion of Fig.\,3, for the sample with in-plane polarisation two distinct relaxation-like features show up, a smaller one reaching $\varepsilon_1$ of about 220 and a more prominent one reaching a CDC for $\varepsilon_2$ of about $2\times 10^3$. Interestingly, for both samples a further increase at low frequencies is revealed, which could be originating from hopping conductivity\cite{Lunkenheimer2010, Jonscher1977}.

%Figure \ref{SFigure1} shows the frequency-dependent dielectric constant (a) and conductivity (b) of an ErMnO$_3$ single crystal oriented with the polarisation perpendicular, i.e. out-of-plane, and parallel, i.e. in-plane, to the attached electrodes. The dielectric data of the out-of-plane sample (closed symbols) were taken from Ref. \cite{Ruff2018APL} and both samples belong to the same batch. 

\begin{figure}[htb]
    \centering
    \includegraphics[width=0.7\linewidth]{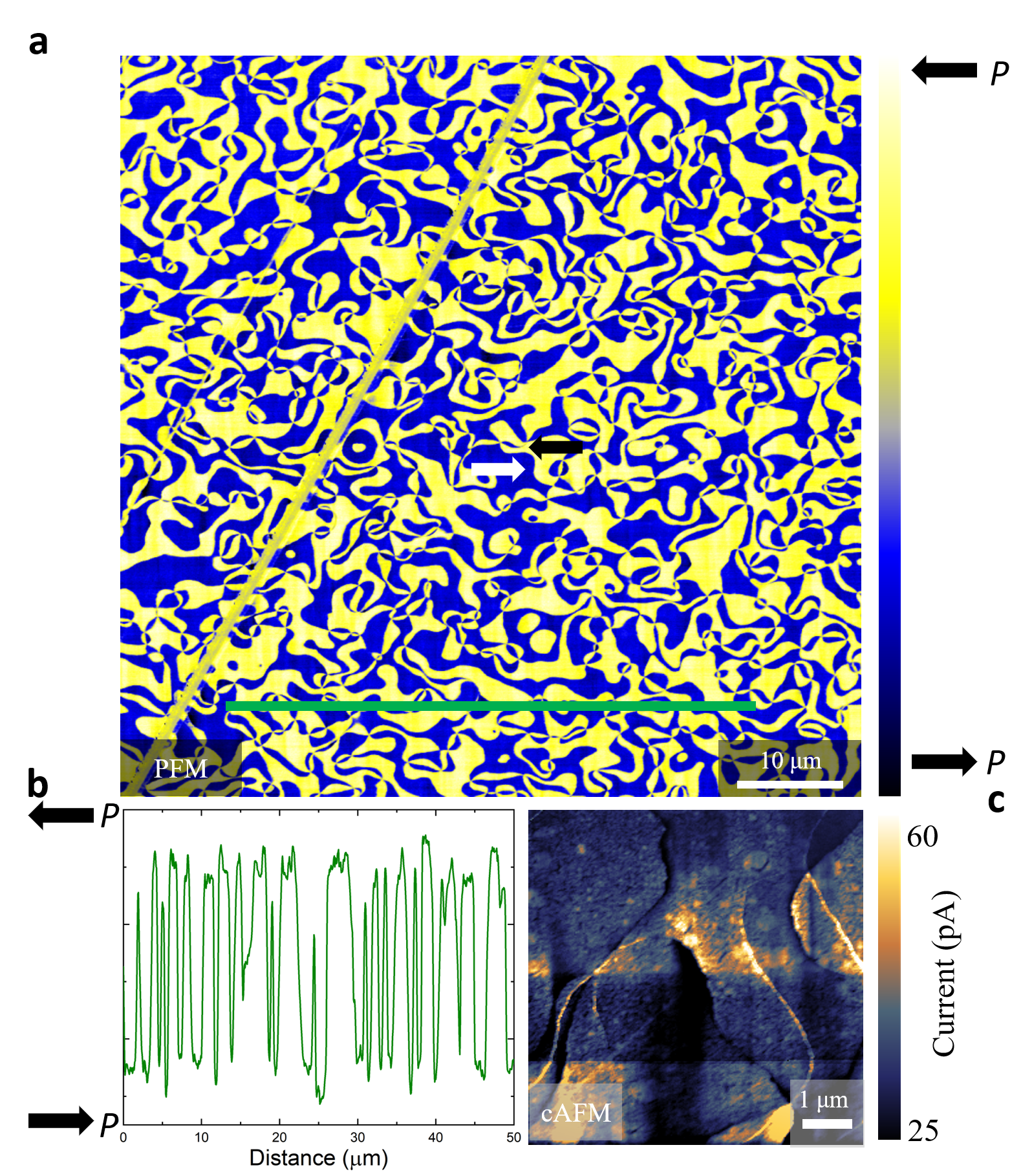}
    \caption{\textbf{a} Calibrated in-plane PFM scan on h-(Er$_{0.99}$,Ca$_{0.01}$)MnO$_3$ where the yellow (blue) represents ferroelectric domains pointing to the left (right). These form the characteristic domain pattern present in hexagonal manganite single crystals, where the ferroelectric $180^{\circ}$ domains come together at sixfold vertex points. \textbf{b} A representative line-profile from \textbf{a} where high values are from domains with \textit{$\leftarrow$P} and low values are from \textit{$\rightarrow$P} domains, used to get an approximate density of DWs per length. \textbf{c} A cAFM image from the area marked be the red box in \textbf{a}. Light colors indicated areas of enhanced conductance (tail-to-tail walls) while dark areas indicate areas of lower conductance (head-to-head walls).}
    \label{SFigure2}
\end{figure}

Figure S\ref{SFigure2} provides the characterization of the ferroelectric domain pattern measured by PFM \textbf{a} and cAFM \textbf{c}. The density of the domain walls with enhanced and reduced conductance in comparison to the bulk is measured by several line-scans. For the DW BLC model we only take into account the number of the DW with reduced conductance, which is for \textbf{b} 0.46 per \textmu m.

\subsection{Quantifying the DW density of ErMnO$_3$}

\begin{figure}[htb]
    \centering
    \includegraphics[width=0.7\linewidth]{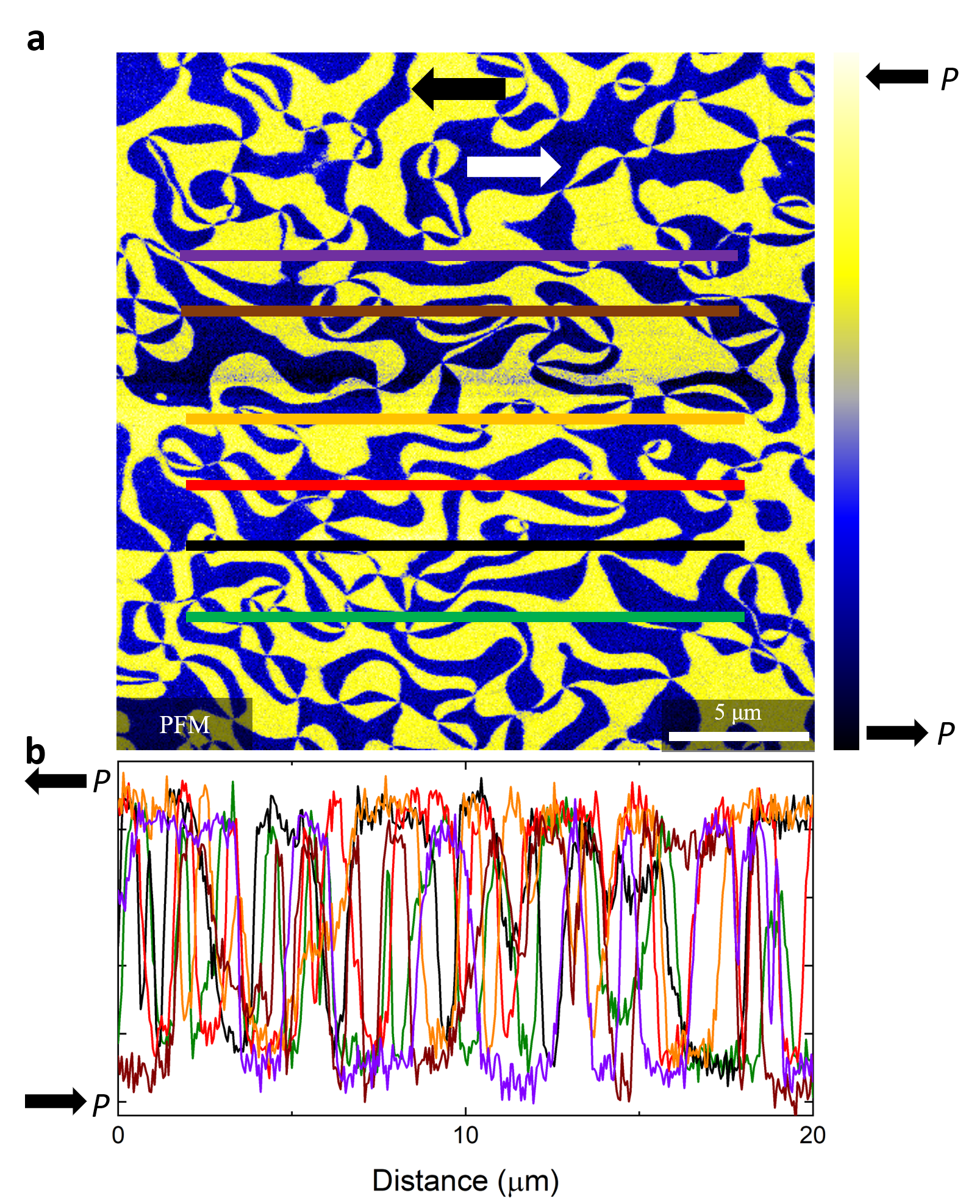}
    \caption{\textbf{a} In-plane PFM scan on h-ErMnO$_3$ with in-plane polarisation (c.f. Fig.\,2). \textbf{b} Line-scans (coloured lines in \textbf{a}) are extracted, from which we deduce an average amount of domain walls with reduced conductance of 10$\pm$1 per 20\,\textmu m. High values are from domains with $\leftarrow$\textit{P} and low values are from $\rightarrow$\textit{P} domains.}
    \label{SFigure3}
\end{figure}

Figure S\ref{SFigure3} shows the PFM measurement of h-ErMnO$_3$ as well as the corresponding lines for the evaluation of the DW density. 

